\documentclass[smallextended]{svjour3} 
\smartqed 
\usepackage{graphicx}
\usepackage{lineno,hyperref}
\usepackage{amssymb}
\usepackage{lineno}
\usepackage[cmex10]{amsmath}
\usepackage[lined,boxed]{algorithm2e}
\usepackage{amsmath}
\usepackage{lineno}
\usepackage{amssymb}
\usepackage{multirow}
\usepackage{caption,setspace}
\usepackage{xspace}
\usepackage{caption}
\usepackage[numbers,sort&compress]{natbib}
\begin{document}

\title{\textit{FusionDeepMF}: A Dual Embedding based Deep Fusion Model for Recommendation}                      
\titlerunning{\textit{FusionDeepMF}: Deep Fusion based Recommendation}        

\author{$^{1, *}$Supriyo Mandal         \and
        $^{2}$Abyayananda Maiti 
}

\authorrunning{Supriyo Mandal         \and
        Abyayananda Maiti} 

\institute{ $^{1}$ Information Profiling and Retrieval, \\
              ZBW-Leibniz Information Centre for Economics,\\
              Düsternbrooker Weg 120, 24105 Kiel, Germany.\\
$^{2}$Department of Computer Science \& Engineering, \\
              Indian Institute of Technology Patna,\\
              Bihta, Bihar, India 801106.\\
              \email{$^{1}$s.mandal@zbw.eu, $^{2}$abyaym@iitp.ac.in.}\\
              ORCID iD: {$^{1}$0000-0002-1310-7291.}\\
              {$^{*}$ corresponding author.}}

\date{}

\maketitle
\begin{abstract}

Traditional Collaborative Filtering ($CF$) based methods are applied to understand the personal preferences of users/customers for items/products from  rating matrix.
Usually, rating matrix is sparse in nature.
 %
%
So there are some improved variants of $CF$ method that apply the increasing amount of side information to handle the sparsity problem.
Only linear kernel or only non-linear kernel is applied in  most of the available recommendation related work to  understand user-item latent feature embeddings from data. 
Only linear kernel or only non-linear kernel is not sufficient to learn complex user-item features from users' side information.
Recently, some researchers have focused on hybrid models that learn some features with non-linear kernel and some other features with linear kernel.
But it is very difficult to understand which features can be learned  accurately with linear kernel or with non-linear kernel.
To overcome this problem,
we propose a novel deep fusion  model named \textit{FusionDeepMF} and 
the novel attempts of this model are i) learning user-item rating matrix and side information through linear and non-linear kernel simultaneously, ii) application of  a tuning-parameter determining the trade-off between the dual embeddings that are generated from  linear and non-linear kernels.
Extensive experiments on  online review datasets  establish that
$FusionDeepMF$ can be remarkably futuristic compared to other baseline approaches.
Empirical evidence also shows that $FusionDeepMF$  achieves better performances compare to linear kernel of Matrix Factorization ($MF$) and non-linear kernel of Multi-layer Perceptron ($MLP$).
%
\keywords{Recommendation System, Deep Neural Network, Matrix Factorization, Reliability score, Review Network.}
\end{abstract}
%
%

\section{Introduction}\label{sec:introduction}

In our daily life,  recommender systems~\cite{history} play a vital role to suggest right products, right social media, preferable news or travel places to the target users in different sectors. 
Using the knowledge of demographic information, past view history and previous purchasing history, recommender systems understand the users' preferences and suggest suitable products or preferable news or close friends to target users~\cite{history,schafer2001commerce}.
An efficient recommendation model can not only understand users' preferable zones but also increases both contentment for users and earnings for companies.
The Content-Based model~\cite{CB1,linden2003amazon} and Collaborative Filtering ($CF$) model~\cite{jacobi2000system,sarwar2001item} are the two important categories of recommender systems. 
The Memory-Based approach~\cite{MB2}  and Model-Based approach~\cite{koren2009matrix} are two types of $CF$. 
Memory-based models generally investigate the customer's preferable items from  rating values given by users on items.
Each entry of a rating matrix is formed based on users' rating values for their purchased items.
But a model-based approach investigates the way users give ratings on items. 
It claims more adequate efficiency than the memory-based approach.

%
Initially, researchers focused only on users' previous rating behavior activities to understand their preferences.
However,  only  previous rating activities are insufficient for rating prediction of an unknown product.
In recent studies, researchers are focusing on  explicit feedback and implicit feedback as users' side information~\cite{weng2004feature}.
Implicit feedback specifies users' activities before purchasing any products.
Users' view activities, user-product interaction, analyzing product image, reading reviews are considered as implicit feedback.
Implicit feedback indicates users' preferences.
Explicit feedback (ratings, reviews) indicates users' activity after purchasing any product~\cite{xiong2018deep}.
It indicates users' satisfaction level.
In~\cite{kim2016convolutional,park2017also,itemcontent} the authors have focused only on implicit feedback.
For explicit feedback, most of the researches consider users' ratings and reviews~\cite{baseline1,liang2016factorization,zheng2019distributed,chen2019user}.
In~\cite{kim2016convolutional,chin2018anr,chen2018neural,w2}, the authors have focused on analyzing review text which leads to a better  rating prediction.
Some researchers have focused on user-product interaction  value based on rating activities~\cite {xue2017deep,dong2017hybrid,he2017neural,fu2018novel}.
There are few works, where the authors have focused on both explicit rating values and implicit feedback~\cite{chen2018matrix,koren2008factorization,li2016exploiting,li2018finite}.
%
%
%
%
%
In~\cite{zhang2017autosvd++,xue2017deep,dong2017hybrid,cheng2018delf,zhang2019deep},  hybrid models are developed.
These models   consider  users' explicit and implicit feedback simultaneously.
%
%
%
%
%
%

It is shown that in review forums of e-commerce sites,  there are some positive reviewers (who always give  positive or good ratings), some negative reviewers (who always give  negative or poor ratings), some reliable reviewers (who give ratings according to the quality of products) and some reviewers are whimsical (whose posted ratings are unpredictable).
User buy products based on their preferable zones but their rating behaviour activities should relate to their characteristics  as potential  reviewers.
We have to understand  whether a user is a positive  or a negative  or a reliable reviewer. 
In most of the papers, when the existing approaches learn latent features of  a user based on  similar behaviours among other users' rating activities or preference degree, they  overlook the characteristics of users.
%
%
%
%

To understand users' characteristic, in this work we consider users'  reliability score as side information.
This reliability score indicates the loyalty and influential power of  users' reviews for their purchased products. 
A user's reliability score varies over different  purchased products.
So, we evaluate this score of a user for a particular product.
Some research works~\cite{moradi2015reliability,o2005trust} have used the term ``reliability''.
But this  score just indicates users'  trust score.
%
%
%
The evaluation of our proposed score  is demonstrated in Section~\ref{sec:3} and the evaluation process is completely different from the existing works.
In~\cite{bobadilla2018reliability,nunez2018recommender}, the researchers propose the concept of reliability based on support for users (the total ratings posted by the user) and support for products (how many ratings are posted for this product). 
This concept is totally different from our evaluation technique.
In \cite{shen2019sentiment}, the authors incorporate the ratings, reviews, and feedback into a probabilistic matrix factorization framework for prediction.
They design users' reliability measure that combines their consistency between ratings and sentiment of reviews and the helpfulness feedback on their reviews.
We have used this method \cite{shen2019sentiment} as a baseline model.
Our reliability concept is different from this baseline work and the concept of learning methodology is also different (discussed in Section~\ref{sec:4}).

%

Based on our experimental data we form a user-product bipartite network,
where all existing side information are coming from the users directly.
The  reliability score is computed from  the  bipartite network between users and products. 
Consequently we classify reliability score as a network feedback score~\cite{mandal2021rating,mandal2021deep}.
It is neither an explicit feedback nor an implicit feedback.
Traditional recommendation systems mainly focus on  reviewers role, their side information as explicit and implicit feedback to predict rating.
In our  work, the raters also play a vital role to notify the reliability of a reviewer's review and help the model to learn the latent factor of the reviewer more accurately to predict rating. 
Most of the previous works ignore  raters' role in recommendation systems.    %


\vspace{4 pt}

\noindent\textbf{Model Architecture:}
Although  users' side information and   selection of features have a crucial role in efficient recommendation, model architecture is another vital part in recommendation.
%
%
%
%
The Matrix Factorization (\textit{MF})~\cite{shi2018heterogeneous,yu2017attributes,ji2016improving,yu2018joint} approach is a popular model-based approach because of its accurate recommendation and high scalability~\cite{scale1,scale2}. 
This model factorizes both products and users by a low-dimensional feature vector space, where every latent vector is treated as a user's feature vector.
A lot of works~\cite {mnih2008probabilistic,WMF,itemcontent,mandal2018explicit,huang2016probabilistic}  are continuing for  accurate recommendation using probabilistic \textit{MF} model. 
The researchers are focusing on users'  implicit~\cite{he2017neural,park2017also} or explicit feedback~\cite{kim2016convolutional,itemcontent} to increase  learning quality of user-product features embeddings.
For some users, linearity of a kernel can model features vectors based on users' side information to predict ratings more accurately.
But for some other users, the linear kernel-based methods may fail to learn the features of users and products~\cite{WMF}.

To overcome the above problem, recently most of the researchers are employing deep neural networks (\textit{DNNs})~\cite{zhang2016collaborative,zheng2017joint,wang2020tdcf}, which use nonlinear kernel  to model latent features based on users' side information.
%
%
In  Deep belief Networks
features are learned automatically~\cite{chen2018neural,he2017neural} (effective feature representation is learned from the dataset).
%
%
Deep learning model achieves its popularity by its scalability for dimension reduction technique.
In~\cite{elkahky2015multi}, the authors use Multi-view \textit{DNNs} to match rich user features to product features.
Another reason for its wide popularity is that \textit{DNNs} are scalable to large datasets.
In~\cite {xue2017deep,dong2017hybrid,he2017neural,fu2018novel}, the authors have  developed deep learning model to learn user-product interaction value based on rating activities.
Due to its fast features learning technique and latent vector dimension reduction technique, deep learning model is also preferred in our research.
However, for some users' non-linear model learn users' latent features very accurately, but for some users non-linear kernel fails~\cite{xue2017deep}.

It is very evident that both linear and nonlinear models can not learn all users' latent features independently or separately.
Recent studies are focusing on hybrid models that learn some features with non-linear kernel and some features with linear kernel.
In \textit{ConvMF}~\cite{kim2017deep}, the authors apply \textit{CNN} on users' posted reviews, products' descriptions and learn users' rating activities using probabilistic matrix factorization. 
In \textit{VMCF}~\cite{park2017also}, the authors apply \textit{CNN} to extract features from pictures of  products and apply \textit{MF} on users' rating activities and ``also-viewed" relationships.
But it is very difficult to understand which features are learned  accurately with linear kernel and which features are learned accurately with non-linear kernel.
From our experiments, we have observed  that both kernels are effective with different set of users.

To overcome this problem,
%
we propose a fusion based model \textit{FusionDeepMF}, where a \textit{MF} based model applies a linear kernel to model the latent features based on users' rating activity  and reliability score   and on the other hand a \textit{MLP} based model uses non-linear kernel to learn users' latent feature vectors. 
In 
concatenation layer of our model,
we combine two separate embeddings, where one embedding is  the output embedding of the \textit{MF} model and another one is  the output embedding of the \textit{MLP} model.
The embedding of the \textit{MF} model indicates user's rating vector which is learned through the \textit{MF} model based on the users' rating activities and  reliability scores. %
Another embedding presents user's rating vector which is learned through the  \textit{MLP} model based on  rating activities and  reliability scores. 
In this layer two separate embeddings can mutually reinforcement each other to learn better rating values.

In this paper, our major contributions are :
\begin{itemize}
    \item We consider reliability score as a network feedback which indicates users' characteristics as reviewers. 
    
    
    \item Introduce \textit{FusionDeepMF} model and two novel characteristics of this model are i) learning user-item rating matrix and side information through linear and non-linear kernel simultaneously, ii) usage of  a tuning-parameter determining the trade-off between the dual embeddings that are generated from linear and non-linear kernels of our model.
%
%


%


\item We extensively evaluate the initialized input parameters of \textit{FusionDeepMF} model through pre-training phase and demonstrate the effectiveness of our proposed model by comparing it with  the state-of-the-art models.
\end{itemize}

The rest of the  paper is categorized as follows:
In section~\ref{sec:2}, we explain our problem and intuition.
Then we introduce our  side information as network feedback in section~\ref{sec:3}.
How we learn users' feature vector embeddings using our model  \textit{FusionDeepMF}, is discussed in section~\ref{sec:4}.
The result of comprehensive experiments on  online review datasets is discussed in section~\ref{sec:5}.
Finally, conclusion and future works are presented in section~\ref{sec:6}.


\section{Problem Formulation}\label{sec:2}

\subsection{Notation}

%
%
%
Suppose, $\mathcal{U}$ = $\lbrace$ $u_{1}$, $u_{2}$, ...., $u_{n}$ $\rbrace$ be the users set, $\mathcal{P}$ = $\lbrace$ $p_{1}$, $p_{2}$, ...., $p_{m}$ $\rbrace$ be the products set, $\mathcal{R}$ = $\lbrace$ $r_{1}$, $r_{2}$, ...., $r_{N}$ $\rbrace$ be the ratings set, $\mathcal{H}$ = $\lbrace$ $rel_{1}$, $rel_{2}$, ...., $rel_{N}$ $\rbrace$ be set of the reliability scores,  where $n$ and $m$ represent  the count of users and  products, respectively.
$N$ represents the count  of both ratings and reliability scores.
First, we build a user-product rating matrix $R$ $\in$ $\mathbb{R}^{n \times m}$ from dataset, where each entry $r_{ij}$ is the rating score given by user $u_{i}$ on product $p_{j}$.
Similarly, we build a user-product reliability score  matrix $H$ $\in$ $\mathbb{R}^{n \times m}$ from  dataset, where each entry $rel_{ij}$ is the reliability score gained by user $u_{i}$ from her review on $p_{j}$.
The range of reliability score $rel_{ij}$ is in (0,1) and  the raw rating $r_{ij}$ is also mapped into  (0,1]  to learn parameters efficiently.
The sigmoid function is chosen to map the inner product of latent  vectors into  [0,1] range.

\subsection{Problem Definition}

 \textbf{Input:} 1) The user-product rating matrix $R$ $\in$ $\mathbb{R}^{n \times m}$ from our dataset where each entry $r_{ij}$ is the rating value given by user $u_{i}$ on product $p_{j}$,
2) The user-product reliability score matrix $H$ $\in$ $\mathbb{R}^{n \times m}$ from our dataset where each entry $rel_{ij}$ is the reliability score of  the rating of user $u_{i}$ on product $p_{j}$.

\noindent\textbf{Output:} The primary goal is to characterize the users' latent feature vectors based on reliability and rating score in a more accurate way, and to give a better interpretation of how the similarity of users' reliability scores affect rating prediction of the unknown products using linear kernel and non-linear kernel.

\section{Network Feedback: users' Side Information}\label{sec:3}

Networks are extensively used in different fields of research as a convenient presentation of patterns of interaction between appropriate staffs.
In research, social network, biological network, ecological network etc. are widely used.
Based on our experimental data, we form a network named \textit{Review Network}.

\vspace{3 pt}
\noindent \textbf{ Review Network: }
Before purchasing any product from e-commerce sites, we check the previous users' opinions regarding that product.
A network of users is formed  depending on reviewing activities of users on different products.
We have named this network as \textit{Review Network} and it is discussed in our previous work~\cite{mandal2018explicit,AMaitiPatent,mandal2020explicit}. 

\subsection{Rating Score }
In e-commerce  sites,  a user posts ratings on purchased product.
Here  $r_{ij}$ is denoted as  rating  of user $u_{i}$ on $p_{j}$.
In this paper, the rating range is  $1$ to $5$  similar to the Amazon.com dataset. 
We consider rating $1$ and $2$ as negating ratings and $3$ to $5$  as positive ratings.
Rating $r_{ij}$ is normalized into  (0,1] by using  function $f(x)=x/5$ (5 is the maximum rating).
%
%
%
%

%
%
%
%

\subsection{Reliability Score} \label{help}

We propose  the reliability score of a user for a specific product.
%
%
The reliability value of a user might vary for the different products the user purchased.
This score of a user for each review is calculated depending on the number of helpful ``yes" votes on the review, how many other users read the review as  most current reviews and top ranking reviews.
%
Here we present the overview of the reliability score in review network.
%

Before buying any product, users can check the previous feedback or reviews regarding that particular product.
From these reviews, user can judge the quality of the product through the satisfaction levels of the previous users.
In most of the e-commerce sites, after each review, sites post a query, ``Was this review helpful to you? (Answer Yes/No)''. 
We define helpfulness as a measure for the validation of a review. 
We normalize the helpfulness $h_{ij}$ of user $u_{i}$ for product $p_{j}$ (i.e., $u_{i}$ wrote review $f_{ij}$ on $p_{j}$) based on how many other users are helpful from the review and the normalized helpfulness score $h_{ij}$ is represented as follows:  
\begin{equation}\label{eq:h}
h_{ij}= \dfrac{l_{ij}}{\sum_{x=1}^{n'}l_{xj}},
\end{equation}
where,
\begin{equation}\label{eq:h1}
l_{ij} = \frac{( \#helpful \ votes \ on \ review \ f_{ij})^{2}}{total \ \# votes \ on \ review \ f_{ij}}.
\end{equation} 
Here, the number of users who buy $p_{j}$, is denoted by $n'$.
We assign more weight to the  particular users who gain more helpful votes.
Thus,  Eq.~\ref{eq:h1} is quadratic in nature.
%
As a special case, if a dataset contains only  the helpful vote information without the number of total votes, then the denominator of Eq.~\ref{eq:h1} will be replaced by the maximum helpful votes gained by any review on product $p_{j}$.

Usually, when a user wants to read the previous reviews for a particular product, the user has two options to choose, $i.e.,$ i) read the top-ranking reviews and ii) read the most recent reviews.
%
%
Helpful votes of the reviews influence us to read.
That is why any user prefers the top-ranking feedback (top-ranking feedbacks are selected based on helpful votes).
%
%
Sometimes users prefer to read both the most recent reviews of a particular product to understand the current quality of the product.
Some users prefer to read the top-ranking reviews and most recent reviews to come to a purchasing decision.
%
%
%

Based on the potential reading activities of the most current reviews we evaluate most recent  score $most_{ij}$ of user $u_{i}$ for product $p_{j}$ and the equation is as follows:
\begin{equation}\label{eq:mi}
c_{ij}= \sum_{s=1}^{n'-i}\dfrac{1}{s^2},
\end{equation}
and the normalized $c_{ij}$ ($i$ value range is 1 to $n'$) is denoted as $most_{ij}$.
\begin{equation}\label{eq:m}
most_{ij}= \dfrac{c_{ij}}{\sum_{x=1}^{n'}c_{xj}},
\end{equation}
where, $most_{ij}  \in (0,1)$ and the denominator is a summation defined by $x$ and $x$=1 to $n'$. 
Here, $n'$ denotes the count of users who posted feedback for $p_{j}$ and $n'-i$ is the  count of users who  buy  $p_{j}$ after $u_{i}$.
It is shown that in Fig.~\ref{fig:review}, $u_{1}$ is the first user who posts feedback for product $p_{1}$.
When $u_{2}$ buys the exact same, she may check $u_{1}$'s feedback as most current review and for $u_{2}$, $u_{1}$'s score = $(1/1^2)$.
%
When $u_{3}$ user buys this product, she may check  $u_{1}$'s feedback as second most current feedback and for $u_{3}$ her  score = $(1/2^2)$ and for $u_{4}$, $u_{1}$'s  score = $(1/3^2)$. 
The score of $u_{1}$ user for product $p_{1}$ = $\sum_{s=1}^{4}\dfrac{1}{s^2}$ = $(1/1^2)$ + $(1/2^2)$ + $(1/3^2)$ + $(1/4^2)$.
%
%
We assign more weight to the most current reviewers regarding the same product.
So the above Eq.~\ref{eq:mi} has  quadratic nature.
%

%
The top ranking feedback is evaluated   depending on  helpfulness score of users' reviews.
Based on the reading activities of the top ranking reviews we evaluate top ranking reviewer score $top_{ij}$ of user $u_{i}$ for product $p_{j}$ and the equation is as follows:
\begin{equation}\label{eq:tt}
q_{ij} =(\dfrac{1}{o_{ij}^2})* (n'-i),
\end{equation}
and the normalized $q_{ij}$ ($i$ value range is 1 to $n'$)  is denotes as $top_{ij}$.
\begin{equation}\label{eq:t}
top_{ij} = \dfrac{q_{ij}}{\sum_{x=1}^{n'} q_{xj}},
\end{equation}
where,  $top_{ij} \in (0,1)$ is the  score of user $u_{i}$ for product $p_{j}$ according to the top ranking reviews.
Here $o_{ij}$ denotes the ranking for user $u_{i}$ for her purchased product $p_{j}$.
The top rank depends on high helpfulness score.
Here, $n'-i$ denotes the count of  users who buy the  product after $i^{th}$ user and read that review.
%
%
We assign more weight to the top ranking reviewers regarding the same product.
So the above Eq.~\ref{eq:tt} is quadratic in nature.

Based on the reading activities of the top ranking reviews and most current reviews, we evaluate total score $d_{ij}$   as follows:
\begin{equation}\label{eq:c}
d_{ij}= \alpha*top_{ij}+(1-\alpha)*most_{ij},
\end{equation}
where $\alpha$ $\in [0,1]$ denotes the  weightage.
Here 0.5 is considered as the weightage, because we want to give the same priority to both $top_{ij}$ and $most_{ij}$.
%
%
The value of $d_{ij}$ $\in (0,1)$.

We define \textbf{reliability score} $rel_{ij}$ of review $f_{ij}$ as the average of the two scores:
\begin{equation}\label{eq:rel2}
rel_{ij}= \dfrac{h_{ij}+d_{ij}}{2},
\end{equation}
where the value of $rel_{ij}$ $\in (0,1)$.
If reliability value $\geq$  0.5 (industry shall decide  the threshold;  here the threshold value is assigned to 0.5), that means the user is a reliable user.
If reliability value $<$ 0.5,  the user is not a reliable one and  rating activity indicates that the user is either a positive or a negative reviewer.
%
%
%





\section{ \textit{FusionDeepMF}: A Methodology using  Reliability Score}\label{sec:4}

In this section, we  explore the effectiveness of users' reliability score  to learn latent feature vectors of users and products  for rating prediction.
Users' latent feature vectors based on the reliability score and rating value indicate their reliability as reviewer, validation of their reviews and rating behavior activities.
%
%
We have already discussed that independently only the linear kernel of a linear model or only the non linear kernel of a non linear model is not  sufficient to learn complex user-product features.
Our experiments establish the fact.
We propose a  model named  \textit{FusionDeepMF}, where MF  uses a linear kernel to model the latent features based on users' reliability score and rating values, and on the other hand MLP model uses a non-linear kernel to learn users' reliability score and rating values for their purchased products.
In the
concatenation layer of our model as represented in Fig.~\ref{fig:MFMLP},
we integrate two separate embeddings $i.e.,$ $\theta^{MF}$ and $\theta^{MLP}$, where
$\theta^{MF}$ is  the final embedding of the \textit{MF} model and %
$\theta^{MLP}$ is  the final embedding of the \textit{MLP} model.
The $\theta^{MF}$ indicates user's latent features vector that is learned through the \textit{MF} model based on the user's rating activities and  reliability scores  for the user's  purchased products. 
The $\theta^{MLP}$ presents user's latent features  vector that is learned through the \textit{MLP} model based on the user's  rating activities, reliability scores  for a particular product.  
In this model two separate embeddings can mutually reinforcement each other to learn a better rating value.
The $\theta^{MF}$ and $\theta^{MLP}$,  these two embeddings are crucial in our model.
How we generate these two embeddings through \textit{MF} and \textit{MLP}, is discussed in  the next subsections.

\subsection{Formation of $\theta^{MF}$ Embedding through \textit{MF} Model}



This model uses linear kernel to learn latent features based on users' rating scores and reliability scores.
Using this model we  investigate the effectiveness of rating prediction based on reliability score. 
The input of the \textit{MF} model is the data of  $u_{i}$ for product $p_{j}$ based on the user's rating score $r_{ij}$ and reliability score $rel_{ij}$.
%
%
The reliability score is evaluated from Eq.~\ref{eq:h} to~\ref{eq:rel2}.
%

\subsubsection{Rating Prediction: Capturing Users' Rating Score}
In this section, we  discuss how we learn users' and products' latent features vectors based on users' rating behaviour activity and predict ratings. 
First, we build a user-product rating matrix $R$ $\in$ $\mathbb{R}^{n \times m}$ from our dataset where each entry $r_{ij}$ is the users' posted rating value given by $u_{i}$ on product  $p_{j}$.
%
%
Singular Value Decomposition ($SVD$) is a popular matrix factorization technique that represents each user and product as a real valued vector of latent features.
We  apply $SVD$\footnote{https://medium.com/@jonathan\_hui/machine-learning-singular-value-decomposition-svd-principal-component-analysis-pca-1d45e885e491} to  the user-product rating matrix $R$ $\in$   $\mathbb{R}^{n \times m}$ and decompose into three matrices $W'$ $\in$ $\mathbb{R}^{K\times n}$, $S_o$ $\in$ $\mathbb{R}^{K\times K}$  and $Z'$ $\in$ $\mathbb{R}^{K \times m}$, where  $S_o$ is a diagonal matrix whose elements (singular values) are equal to the root of the positive eigenvalues of $RR^T$ or $R^TR$.
Then after observing the performance of our model, we  decide to  generate $W$ $\in$ $\mathbb{R}^{K\times n}$ = $[W'^{T} *  S_o^{1/2}]^T$  
and similarly,
$Z$ $\in$ $\mathbb{R}^{K\times m}$ = $[S_o^{1/2}* Z']$.
%
In the two matrices $W$ $\in$ $\mathbb{R}^{K\times n}$ and $Z$ $\in$ $\mathbb{R}^{K \times m}$, where $w_{i}$ $\in$ ${W}^{K\times n}$ ( $w_{i}$ is the $i^{th}$ column of ${W}^{K\times n}$ ) is the user-specific latent feature vector of user $u_{i}$ and 
$z_{j}$ $\in$ ${Z}^{K\times m}$ ($z_{j}$ is the $j^{th}$ column of ${Z}^{K\times m}$) is the product-specific latent feature vector of product $p_{j}$.
By applying low rank approximation method to the observed user-product rating matrix, unknown ratings by $W^TZ$ are predicted.
Together, the $w_{i}^Tz_{j}$ indicates how user $u_{i}$ rates on product $p_{j}$.
Formally, the feature matrices $W$ and $Z$ can be learned by minimizing a loss function as follows:
\begin{equation}
\begin{split}
\varphi = \min\limits_{W,Z}   \sum_{i=1}^{n}\sum_{j=1}^{m}
I_{ij}^{R}(r_{ij}-g(w_{i}^T z_{j}))^2
+\lambda\bigg(\sum_{i=1}^{n} n_{w_{i}}\parallel w_{i} \parallel_{F}^{2}
+\sum_{j=1}^{m} n_{z_{j}}\parallel z_{j} \parallel_{F}^{2} \bigg),
\end{split}
\label{eq:obj1}
\end{equation}

%
%
where, $\|$.$\|_{F}$ represents the Frobenius norm and 
$g(.)$ indicates Sigmoid function.
The term $\lambda$(.)  is introduced to avoid over-fitting. 
Here, $I_{ij}^{R}$ = 1 if $u_{i}$ rates on product $p_{j}$, otherwise $I_{ij}^{R}$ = 0. 
$n_{w_{i}}$ denotes the total count of ratings, that is given by  $u_{i}$   and $n_{z_{j}}$ is the total count of users who rate on product $p_{j}$.

\subsubsection{Factorization of Reliability Score  Matrix}

In this section, we  discuss how we learn users' and products' latent features vectors based on users' reliability scores.
Here, we build a user-product reliability score based matrix $H$ $\in$ $\mathbb{R}^{n \times m}$ from our dataset where each entry $rel_{ij}$ is the reliability score gained by  $u_{i}$ from the user's review on $p_{j}$.
%
%
We  follow the decomposition technique of our rating matrix  and similarly we decompose  the user-product reliability matrix $H$ $\in$ $\mathbb{R}^{n \times m}$  into two matrices $E$ $\in$ $\mathbb{R}^{K\times n}$ and $F$ $\in$ $\mathbb{R}^{K \times m}$, where $e_{i}$ $\in$ ${E}^{K\times n}$ ( $e_{i}$ is the $i^{th}$ column of ${E}^{K\times n}$ ) is the user-specific latent feature vector of $u_{i}$ based on reliability score and
$f_{j}$ $\in$ ${F}^{K\times m}$ ($f_{j}$ is the $j^{th}$ column of ${F}^{K\times m}$) is the product-specific latent feature vector of  $p_{j}$.
By applying low rank approximation method to the observed reliability matrix, we can recover unknown reliability scores by $E^TF$.
Together, the $e_{i}^Tf_{j}$ indicates how user $u_{i}$'s rating gains reliability score on  $p_{j}$.
Formally, the feature matrices $E$ and $F$ can be learned by minimizing a loss function as follows: 
\begin{equation}
\begin{split}
\varphi = \min\limits_{E,F}   \sum_{i=1}^{n}\sum_{j=1}^{m}
I_{ij}^{rel}(rel_{ij}-g(e_{i}^T f_{j}))^2 
+\lambda\bigg(\sum_{i=1}^{n} n_{e_{i}}\parallel e_{i} \parallel_{F}^{2}
+\sum_{j=1}^{m} n_{f_{j}}\parallel f_{j} \parallel_{F}^{2} \bigg),
\end{split}
\label{eq:rel}
\end{equation} 

where,  $I_{ij}^{rel}$ = 1 if user $u_{i}$ gains reliability score on product $p_{j}$, otherwise $I_{ij}^{rel}$ = 0. 
Here $n_{e_{i}}$ denotes the count of products for which   $u_{i}$ gains reliability score  and $n_{f_{j}}$ is the total count of users who gain  score on  $p_{j}$.

\subsubsection{Rating Prediction: Capturing Users' Reliability Score}
In this section, we  discuss how we predict rating based on users' reliability scores.
In our model, matrix $E$ is the latent space commonly shared by $R$ and $H$, meaning the user feature matrix based on rating $W$ is approximated by user feature matrix based on reliability score $E$.
Here each vector $e_{i}$ characterizes two aspects at the same time: how user $u_{i}$'s review  gains reliability score  for products and how the same user posts ratings on products.
Together, the $e_{i}^{T}z_{j}$ indicates how we predict  $u_{i}$'s rating on product $p_{j}$ based on $u_{i}$'s reliability score similarity with the other users, where $e_{i}$ indicates how user $u_{i}$ gains reliability score for all products and $z_{j}$ indicates the all users' rating activity on  $p_{j}$.
It is the approximation of original rating score $r_{ij}$.
Therefore, we can learn the feature matrices $E,Z$ simultaneously by minimizing the following objective function:  
\begin{equation}
\begin{split}
\varphi = \min\limits_{E,Z}   \sum_{i=1}^{n}\sum_{j=1}^{m}
I_{ij}^{R}(r_{ij}-g(e_{i}^T z_{j}))^2
+\sum_{i=1}^{n}\sum_{j=1}^{m}
I_{ij}^{rel}(rel_{ij}-g(e_{i}^T f_{j}))^2\\
+\lambda\bigg(\sum_{i=1}^{n} n_{e_{i}}\parallel e_{i} \parallel_{F}^{2}
+\sum_{j=1}^{m} n_{z_{j}}\parallel z_{j} \parallel_{F}^{2} 
+\sum_{j=1}^{m} n_{f_{j}}\parallel f_{j} \parallel_{F}^{2} \bigg).
\end{split}
\label{eq:obj}
\end{equation}

In this way, we can integrate two types of data sources and obtain latent spaces $E$ and $Z$, that can work together to give exact prediction.

\subsubsection{ Optimization Technique for \textit{MF} model}
%
%
We adopt a popular method, mini-batch Adaptive Moment Estimation (Adam)~\cite{kingma2014adam}\footnote {We  apply Algorithm 1 in~\cite{kingma2014adam}}, which tunes the learning rate for each parameters by performing smaller updates for frequent and larger updates for infrequent parameters.
Mini-batch Adam optimizer is a popular method for deep learning model.
Our model is the fusion of \textit{MF} and \textit{MLP}.
So, we adopt a common optimizer as mini-batch Adam for \textit{FusionDeepMF}, pre-training phase of \textit{MF} and \textit{MLP}.
The gradient of $\varphi$ (Eq.~\ref{eq:obj1}) with respect to $w_{i}$ is evaluated as described in~\cite{kingma2014adam} to optimize the objective function.  
%
%
Similarly, we evaluate $\frac{\partial \varphi  }{\partial z_{j}}$ from Eq.~\ref{eq:obj1}.
The value of 
$\frac{\partial \varphi  }{\partial e_{i}}$,
$\frac{\partial \varphi  }{\partial f_{j}}$
and 
$\frac{\partial \varphi  }{\partial z_{j}}$ are also evaluated from Eq.~\ref{eq:obj}.
%
and update the parameters.
%
%
After independently training of Eq.~\ref{eq:obj1} and~\ref{eq:obj}, one can obtain two sets of feature matrices.
Let $w_{i}^{rMF}$ and $z_{j}^{rMF}$ be the user and product-specific vector, respectively learned based on users' rating activities.
Let $e_{i}^{vMF}$ and $z_{j}^{vMF}$ be the user and product-specific vector, respectively learned based on users'  reliability scores.
%
After convergent we get final updated values.
%
%
Here,  $w_{i}^{rMF}, z_{j}^{rMF}, z_{j}^{vMF}, e_{i}^{vMF}$ are crucial to generate $\theta^{MF}$ embedding for user $u_{i}$, that represents the user's  feature based on her rating activity and  reliability score.



%

\subsubsection{Learning of $\theta^{MF}$ Embedding}

\begin{figure}[hbtp]
\centering
\includegraphics[scale=.6]{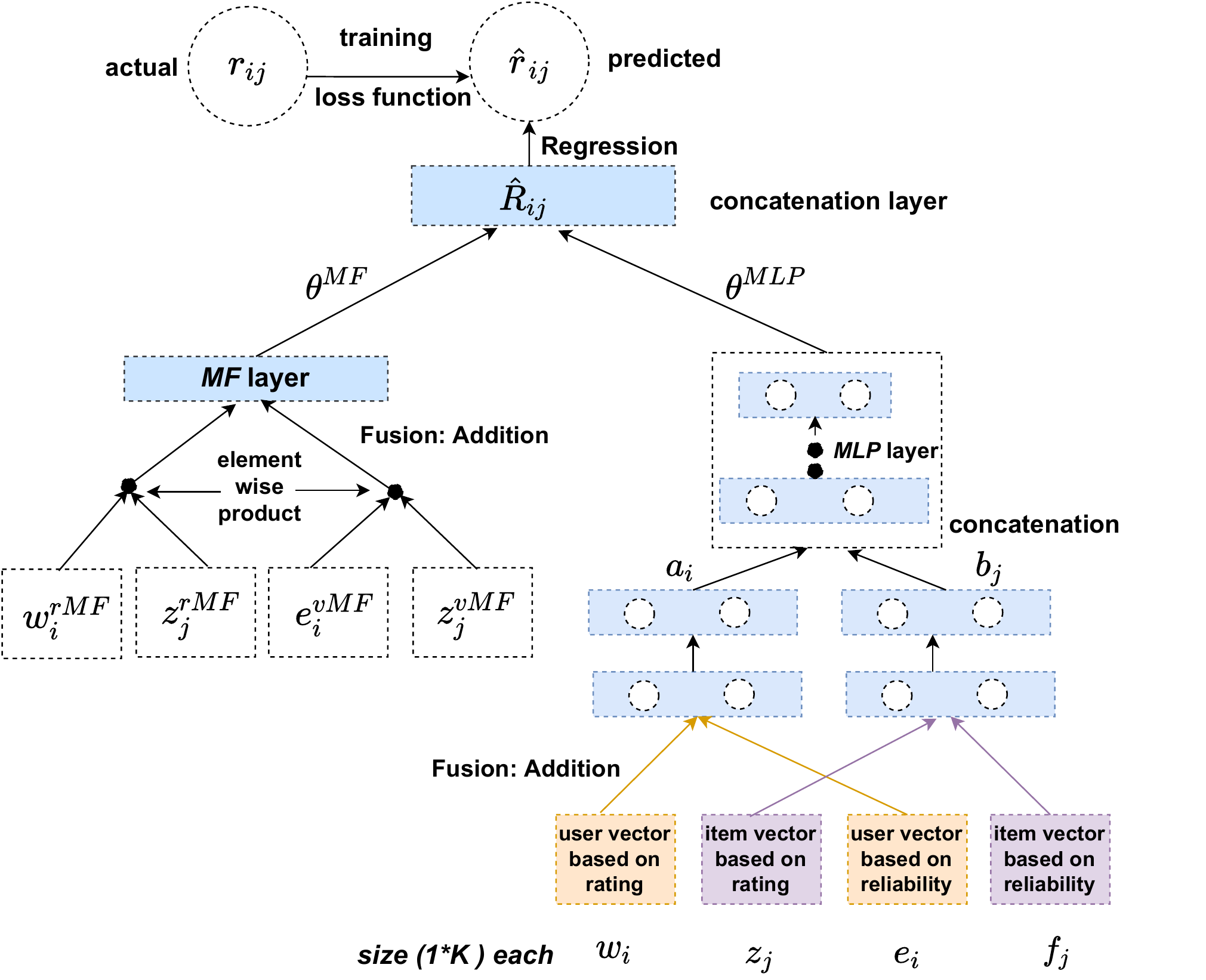}
\caption{\textit{FusionDeepMF: A deep fusion  model architecture with the size of embedding $K$ = 256, where optimal user's feature embeddings of linear model and non-linear model mutually reinforcement each other for a better prediction.}}
\label{fig:MFMLP}
\end{figure}

We generate $\theta^{MF}$ embedding to represent users' latent features vectors based on rating behaviour activities and reliability scores.
%
%
%
In \textit{MF} layer as shown in Fig.~\ref{fig:MFMLP}, we perform addition of two output vector ($K$ dimensional latent features vector each) of $w_{i}^{rMF}\odot z_{j}^{rMF}$ and $e_{i}^{vMF}\odot z_{j}^{vMF}$.
After observing the performances of different fusion methods such as concatenation, addition, or element wise product in our model, addition method is preferred for combining two output vector.
We evaluate $\theta^{MF}$ embedding as follows:

\begin{equation} 
\theta^{MF} =\bigg[ W_{MF1} [w_{i}^{rMF}\odot z_{j}^{rMF}] \ + \ W_{MF2}[e_{i}^{vMF}\odot z_{j}^{vMF}] \ \bigg],
\label{eq:embedding1}
\end{equation}

where, $W_{MF1}$, $W_{MF2}$ denote the weight matrices of the input of \textit{MF} layer. 
Here, $\odot$ indicates element wise product.
The $\theta^{MF}$ indicates user's feature vector that is learned through \textit{MF} model based on the user's rating activities and reliability scores on her purchased products.

%

\subsection{Formation of $\theta^{MLP}$ embedding through \textit{MLP} Model}\label{mlp}
%
%
The \textit{MF} model uses linear kernel to model latent feature embedding $\theta^{MF}$ based on users' rating activities and  reliability scores. 
Our observation is that for some users, the \textit{MF} model uses linear kernel to model latent features very accurately and for some users \textit{MF} model fails. 
So in our model \textit{FusionDeepMF} as shown in Fig.~\ref{fig:MFMLP}, we have generated $\theta^{MLP}$ embedding based on users' rating activities and reliability score through a  deep learning  technique, that uses non-linear kernel to model latent features.
Using this model we investigate the effectiveness of rating prediction based on reliability score through linear and non-linear kernel.
In this section, we discuss a  Multilayer Perceptron Model (\textit{MLP}), that generates $\theta^{MLP}$ embedding. 
%
%

  
\subsubsection{The Input and Embedding Layer of \textit{MLP} model}

The input of the \textit{MLP} model is the same as the initial input of the \textit{MF} model, where  the data of user $u_{i}$ for product $p_{j}$ is based on the user's rating score $r_{ij}$ and reliability score $rel_{ij}$.
%
%
The rating and reliability score are evaluated from Eq.~\ref{eq:h} to~\ref{eq:rel2}.
Suppose, $\mathcal{U}$ = $\lbrace$ $u_{1}$, $u_{2}$, ...., $u_{n}$ $\rbrace$ be the users set, $\mathcal{P}$ = $\lbrace$ $p_{1}$, $p_{2}$, ...., $p_{m}$ $\rbrace$ be the products set, $\mathcal{R}$ = $\lbrace$ $r_{1}$, $r_{2}$, ...., $r_{N}$ $\rbrace$ be the ratings set, $\mathcal{H}$ = $\lbrace$ $rel_{1}$, $rel_{2}$, ...., $rel_{N}$ $\rbrace$ be the set of reliability scores,  where $n$ and $m$ represent the count  of users and  products, respectively.
$N$ denotes the count of ratings and reliability score  in our  dataset.
%
%
%
%
Embedding layer is fully connected.
The dimension of $w_{i}$, $z_{j}$, $e_{i}$, $f_{j}$ is (1 $\times$ $K$),  and each vector is the feature vector of user $u_{i}$ and product $p_{j}$ based on rating, reliability score  and  these vectors are passed to the next layer which is the fusion layer.

\subsubsection{Fusion Layer}
In this layer, the embedded features of users and products based on ratings and  reliability score  are combined for better presentation of  learning.
%
%
After observing the performances of different fusion methods such as concatenation, addition, or element-wise product in our model, 
addition fusion method is preferred for combining users' vectors and products' vectors.
After fusion a vector represents a user's features based on rating and reliability scores and another vector indicates an product's features.
After the addition step, we use a fully-connected neural layer and ReLU activation function  is applied directly for more effective performance and it is also perceived in our experiment.
The outputs of the Fusion layer are $a_{i}$ $\in$  $\mathbb{R}^{K}$ and  $b_{j}$ $\in$   $\mathbb{R}^{K}$ which are the representation vector of user $u_{i}$ and product $p_{j}$ (learned from fusion part), respectively.

\subsubsection{MLP Layer}

Our model acquires two pathways to design users and products latent vectors combining rating and  reliability score.
In the multi-modal deep learning work \cite{multimodal}, this design has been popularly followed.
For designing a deep neural model, simply vectors concatenation are not enough to extract user-product activity from user and product latent features.
To overcome this problem, we  add hidden layers on the concatenated vector and this structure 
can furnish the model with flexibility and non-linearity to understand the user's  latent features vector based on $a_{i}$ and $b_{j}$. 
To obtain  the feature vectors we have performed concatenation, where  $V=[a_{i} ; b_{j}]$.
It is passed into fully connected layers as follows:
\begin{equation}
\begin{split}
\theta^{MLP}=o_{L}(W_{L}(o_{L-1}(W_{L-1}....o_{1}(W_{1}V+b_{1})) 
+b_{L-1})+b_{L}),
\end{split}
\label{eq:mlp}
\end{equation}
where $L$ presents the count of hidden layers.
$W_{L}$, $b_{L}$ and $o_{L}$ present the weight matrix, bias vector, activation function for $L^{th}$ layer's perceptron, respectively.
$\theta^{MLP}$ indicates user's embedding that is learned through \textit{MLP} model based on the user's rating activities, reliability scores.
We  choose ReLU activation function.
In our investigation  the performance of ReLU  is better than tanh and sigmoid.

%
To learn  features of data accurately \cite{he2016deep}, we develop the tower structure of \textit{MLP} layers where the small number of hidden units for higher layers are employed as shown in Figure \ref{fig:MFMLP}.
Our model's performance with different number of layers has been examined and after that,  we  apply four hidden layers.
%
%

\subsection{Fusion between $\theta^{MF}$ and $\theta^{MLP}$ embedding}


To provide more flexibility, we allow pre-training of \textit{MF} and \textit{MLP} to learn in  separate way and combine the  optimal output embeddings $\theta^{MF}$ and $\theta^{MLP}$ in the concatenation layer as shown in Fig.~\ref{fig:MFMLP}.
The formulation of this concatenation is as follows:
%
%
   %
   %
\begin{equation}
\begin{split}
\hat R_{ij}= \bigg(W_{h}\big[\theta^{MF}  ; \ \theta^{MLP}\big ]\bigg),
\end{split}
\label{eq:weight}
\end{equation}

where $W_{h}$  denotes edge weight of the concatenation  layer.
The derivative of the model $w.r.t$ each model parameter can be calculated with standard back-propagation.




\subsection{ Pre-training}
Due to the non-convexity of the objective function of \textit{FusionDeepMF} model, gradient based optimization methods  find only locally optimal solutions.
For the convergence and performance of  \textit{FusionDeepMF} model, initialization plays an important role.
Since our model is an ensemble of \textit{MF} and \textit{MLP}, we decide to initialize \textit{FusionDeepMF} model  using the pre-trained models of \textit{MF}  and \textit{MLP}.

\vspace{5 pt}

\noindent \textbf{Pre-training of \textit{MF} Model:}
We have evaluated predicted rating $\hat r_{ij}$, which does not follow the well-known practice as 
%
    $\hat{r}_{ij}= g \bigg(\dfrac{(w_{i}^{rMF})^{T}z_{j}^{rMF}+ (e_{i}^{vMF})^{T}z_{j}^{vMF}}{2}\bigg)*r_{max}$,
%
because in our observation, it is found that for some users two latent factors $i.e.$, $w_{i}^{rMF}$, $z_{j}^{rMF}$  claim more weightage values out of four latent factors.
For some users,  $e_{i}^{vMF}$, $z_{j}^{vMF}$ two latent factors claim more weightage values.
So, in our work, we propose a different strategy where
%
%
%
in \textit{MF} layer we perform addition of the two output vectors of $w_{i}^{rMF}\odot z_{j}^{rMF}$  and $e_{i}^{vMF}\odot z_{j}^{vMF}$. 
%
%
We evaluate the $\theta^{MF}$ embedding as mentioned in Eq.~\ref{eq:embedding1}.
%

%
This edge-weightage matrix is learned by standard back-propagation technique with mini-batch Adam optimizer and $MAE$ as the loss function with original rating value $r_{ij}$.
$\theta^{MF}$ indicates user's feature vector that is learned through \textit{MF} model based on the user's rating activities, reliability scores.
The final predicted user's latent  vector  $\hat R_{ij}$  is evaluated as follows:
\begin{equation}
\begin{split}
\hat R_{ij}=\bigg((W_{h}^{MF})^{T}\big[\theta^{MF}\big ]\bigg),
\end{split}
\label{eq:weight1}
\end{equation}

where, $W_{h}^{MF}$ is the edge weights of the output layer.
The predicted rating ($\hat r_{ij}$) is obtained via a regression layer: 
\begin{equation}
\begin{split}
\hat r_{ij}=W_{m}\hat R_{ij}+b_{m},
       \end{split}
\end{equation}
where, $W_{m}$ and $b_{m}$ are the weight matrix and bias vector, respectively.
%
%
Each output of our training dataset of our model is trained with the original target  output ($r_{ij}$) with the standard back-propagation technique.
%
%
For our model we use the mini-batch $Adam$ optimizer and $MAE$ as the loss function after experimental observation.

\vspace{5 pt}
\noindent \textbf{Pre-training of \textit{MLP} Model:}
The pre-training of \textit{MLP} model is same as discussed in section~\ref{mlp}.
After evaluating $\theta^{MLP}$ (output of the MLP layer) using the same Eq.~\ref{eq:mlp}, we generate final predicted user's latent vector  
$\hat R_{ij}$  as follows:
\begin{equation}
\begin{split}
\hat R_{ij}=\bigg((W_{h}^{MLP})^{T}\big[\theta^{MLP}\big ]\bigg),
\end{split}
\label{eq:weight2}
\end{equation}

where, $W_{h}^{MLP}$ is the edge weights of the output layer.
The predicted rating ($\hat r_{ij}$) is obtained via a regression layer: 
\begin{equation}
\begin{split}
\hat r_{ij}=W_{mlp}\hat R_{ij}+b_{mlp},
       \end{split}
\end{equation}
where, $W_{mlp}$ and $b_{mlp}$ are the weight matrix and bias vector, respectively.
%
%
Each output of our training dataset of our model is trained with the original target  output ($r_{ij}$) with the standard back-propagation technique.
%
%

We first train \textit{MF} and \textit{MLP} until convergence.
We then use the models parameters as the initialization for the parameters of $FusinDeepMF$.
For Eq.~\ref{eq:weight} the edge weight matrix $W_{h}$ is denoted as follows:
\begin{equation}\label{tuning}
W_{h} \longleftarrow \bigg [\gamma W_{h}^{MF} \ ; (1-\gamma) W_{h}^{MLP} \bigg ],
\end{equation}

where, $W_{h}^{MF}$ and $W_{h}^{MLP}$ denote the weight vector of the pre-trained models \textit{MF} and \textit{MLP}.
The tuning-parameter $\gamma$ learns the trade-off between $\theta^{MF}$ and $\theta^{MLP}$ two embeddings.

\subsection{Regression and Rating Prediction of \textit{FusionDeepMF} Model}  
The predicted rating ($\hat r_{ij}$) of \textit{FusionDeepMF} Model is obtained via a regression layer: 
\begin{equation}
\begin{split}
\hat r_{ij}=W_{re}\hat R_{ij}+b_{re},
       \end{split}
\end{equation}
where, $\hat R_{ij}$ is the output of Concatenation Layer (Eq.~\ref{eq:weight}) of \textit{FusionDeepMF} Model.  
$W_{re}$ and $b_{re}$ are the weight matrix and bias vector, respectively.
%
%
Each output of our training dataset of our model is trained with the original target  output ($r_{ij}$) with the standard back-propagation technique.
%
%
For our model we use the mini-batch $Adam$ optimizer and $MAE$ as the loss function after the experimental observation.

\subsection{Complexity Analysis}\label{timec}

%
The computational time for the  objective function (Eq.~\ref{eq:obj}) is $\mathcal{O}(tK( \left|\Omega \right| + \left| \Psi \right|$)),
where $t$ represents the iteration counts, $K$ presents the dimension of feature vectors, $\Omega$ and  $\Psi$  present the count of available rating values and  reliability score, respectively.
The computing time of gradients  against $W$, $Z$, $E$, $F$ are
$\mathcal{O}$(tK$\left|\Omega \right|$),   $\mathcal{O}$(tK$\left|\Omega \right|$),  $\mathcal{O}$(tK$\left|\Psi \right|$),  $\mathcal{O}$(tK$\left|\Psi \right|$),  respectively.
Therefore, total time complexity of the \textit{MF} model is $\mathcal{O} (tK( \left|\Omega \right| + \left| \Psi \right|$)).

%

The computing time of the \textit{MLP} model is evaluated based on  the computing time of  the objective function of this model
for rating score, reliability score and its gradients against $w_{i}$, $z_{j}$, $e_{i}$, $f_{j}$.
We also  consider the computing time of  the hidden layers and  the number of  iterations. 
Based on the learning of all parameters, the total computing time of the \textit{MLP} model is 
$\mathcal{O} \bigg(t\big(K(\left|\Omega \right| + \left|\Psi\right|) + \sum_{i=1}^{L} T_{l-1} T_{l}\big)\bigg)$.
The additional time $\sum_{i=1}^{L}$ $T_{l-1} T_{l}$, is caused by the hidden layers.
Here $L$ is the number of hidden layers.
For the $l^{th}$  hidden layer, the computing time is $\mathcal{O}$($T_{l-1} T_{l}$).

The total computational time of our model \textit{FusionDeepMF} is $\mathcal{O} \bigg(t\big(K(\left|\Omega \right| + \left|\Psi\right|) + 
\sum_{i=1}^{L} T_{l-1} T_{l}\big)\bigg)$ +  $\mathcal{O} (tK( \left|\Omega \right| + \left| \Psi \right|$)) 
$\approx$  $\mathcal{O} \bigg(t\big(K(\left|\Omega \right| + \left|\Psi\right|) + 
\sum_{i=1}^{L} T_{l-1} T_{l}\big)\bigg)$.
In order to maintain efficiency and accuracy, we  pre-train the model to reduce our model time complexity to  $\mathcal{O} (tK( \left|\Omega \right| + \left| \Psi \right|$)). 
With the idea of pre-training, we put extra time into pre-training without affecting
our framework and this optimization process scales linearly with size of the given data.
\section{Experimental Result}\label{sec:5}
For analyzing performance of \textit{FusionDeepMF} model, we consider the  Amazon.com online review dataset~\cite{he2016ups,mcauley2015image}\footnote{http://jmcauley.ucsd.edu/data/amazon/ \\  https://nijianmo.github.io/amazon/}  on electronics, movies, TV, video etc. 
Data statistics are presented in Table \ref{tab:datast}, where the $1^{st}$ column indicates different category of  data.
Here,  the $2^{nd}$, $3^{rd}$ and $4^{th}$ column indicate the count of users (unique), the count of  products (unique) and the count of users' posted reviews, respectively.
%
%
%
The Amazon.com data contains:
``$reviewerID$'', ``$asin$'', ``$reviewerName$'', ``$helpful$'', ``$reviewText$'',``$overall$'', ``$summary$'',
 ``$unixReviewTime$'' ,``$reviewTime$''.

An example  of  the  dataset is as follows: \\
\textit{\{\\
  ``reviewerID": ``A2SUAM1J3GNN3B",\\
  ``asin": ``0000013714",\\
  ``reviewerName": ``J. McDonald",\\
  ``helpful": [3, 5],\\
  ``reviewText": ``I bought this ..... Great purchase though!",\\
  ``overall": 3.0,\\
  ``summary": ``Heavenly Highway Hymns",\\
  ``unixReviewTime": 126472000,\\
  ``reviewTime": ``09 13, 2009".\\
\}\\}
In this dataset, ``$asin$" presents product id and  ``$overall$" presents rating.
%
Helpfulness score is evaluated from ``$helpful$" entry and ``helpful":[3, 5] indicates three users are helpful from the review  and other two users feel it is not helpful.
We identify the brand name from ``$reviewText$" field of the review.
Users' view-products and pictures of products are collected from~\cite{he2016ups,mcauley2015image}.
It has  ``$asin$'', ``$title$'', ``$price$'', ``$imUrl$'', ``$related$'',``$salesRank$'', ``$brand$'', ``$category$'' entries. 
Here ``$related$'' entry has information about user id(s) who only view products but not buy and the user id(s) who purchase products after viewing products.
The pictures of products are  collected from the ``$imUrl$'' entry.
%
%
  
\begin{table*}[hbtp]
\centering 
\caption{ Data Statistics} 
\scalebox{0.86}{
\begin{tabular}{|c| c| c |c|}
\cline{1-4}
\multirow{1}{*}{\textbf{Dataset}} &\multirow{1}{*}{\textbf{\# users}} & \multirow{1}{*}{\textbf{\# products}}\multirow{1}{*}& {\textbf{\# reviews/ ratings}}
\\ \cline{1-4}
Electronics & 811,034 & 82,067 & 1,241,778 \\ 

Books & 2,588,991 & 929,264 & 12,886,488 \\ 

Music & 1,134,684 & 556,814 & 6,396,350  \\ 

Movies \& TV & 1,224,267 & 212,836 & 7,850,072  \\ 
Home \& Kitchen & 644,509 & 79,006 & 991,794  \\ 
Amazon Instant Video & 312,930 & 22,204 & 717,651  \\ \cline{1-4}
\end{tabular} 
}
\label{tab:datast}
\end{table*}

\subsection{Baseline methods}
In our model, we use reliability score as users' side information.
To investigate the effectiveness of our model with  considered side information, we choose some popular baselines, which are developed based on different side information and model architecture as shown in Table~\ref{tab:pro}.
%
To assess the performance of our methodology, the baselines
are applied on  Amazon.com dataset. 
For implementation purpose we follow the same experimental settings as mentioned in the baseline models~\cite{koren2008factorization,kim2017deep,park2017also,xue2017deep,dong2017hybrid,fu2018novel,chen2019user,shen2019sentiment}.
%
The baseline models are as follows:
%
%
%

\textbf{i) SVD++~\cite{koren2008factorization}:} Both implicit feedback and explicit feedback are integrated in  this model.
Implicit feedback is decided from available rating values.
If a user buys a product and rates it, that meas he likes the product and positive implicit feedback is generated.
Similarly if the  user does not buy a product, that means he does not like the product and negative implicit feedback is derived.
The users' rating activities are treated as explicit feedback.

\textbf{ii) ConvMF~\cite{kim2017deep}:}  Matrix factorization  integrates
convolutional neural network (CNN) into probabilistic matrix factorization (PMF) that captures contextual information of documents to learn users' preference areas.
This model focuses on explicit feedback.
A document latent vector from proposed CNN is used as the mean of Gaussian noise of an product.
The authors propose a model that plays an important role as a bridge between PMF and CNN that helps to analyze both description documents and ratings.

 \textbf{iii) VMCF~\cite{park2017also}:}
This approach integrates  the  product pictures and ``also-viewed" product information for rating prediction.
The authors use CNN to extract visual features.
Then the embedded visual features and embedded ``also-viewed" products are used in proposed PMF model for better rating prediction.

\textbf{iv) DMF~\cite {xue2017deep}:} In Deep Matrix Factorization Model, the authors construct a  matrix with
explicit ratings and non-preference implicit response.
Meanwhile,
the authors mark a zero if the rating is unknown, which is named as
non-preference implicit feedback in this paper.
If there is any interaction between user $u_{i}$ and product $p_{j}$, then it is denoted as explicit rating value where in~\cite{he2017neural} it is denoted by 1. 
In this paper, the predicted interaction value of a user for a new product indicates her rating value on this product.
The model~\cite{xue2017deep} performs better than the popular model~\cite{he2017neural} when user-product interaction feature is considered based on rating activities.
So, we choose \cite{xue2017deep} as a baseline model.

\textbf{v) HCFDS~\cite{dong2017hybrid}:} The authors  utilize advances in learning effective representations in deep learning, and propose a hybrid
model which jointly performs deep users and products’ latent
factors learning from side information and collaborative filtering from the rating matrix.
They use users' rating activities and additional side information of users to learn user-product latent features embeddings.

\textbf{vi) CM+RIM~\cite{fu2018novel}:} In Novel Deep Learning-Based Collaborative
Filtering Model, the authors first capture the overall co-occurrence and  local
co-occurrence given by certain rating using Constraint Model (CM) and Rating Independent Model (RIM), respectively.
They propose a multiviews neural networks
relying on user-product interaction. 
Moreover,
both CM or RIM are fed into the multiviews neural networks
to further enhance the performance of rating prediction.


\textbf{vii) SBMF+R~\cite{shen2019sentiment}:} The authors develop a sentiment
analysis approach using a new star-based dictionary construction technique to obtain the sentiment score.
They design a user reliability measure that combines user consistency and the feedback on reviews.
They also incorporate the ratings, reviews, and feedback into a probabilistic matrix factorization framework for prediction.

\textbf{viii) DBNSA~\cite{chen2019user}:} The authors propose
a deep learning model to process user comments and to generate
a possible user rating for user recommendations. 
First, the system
uses sentiment analysis to create a feature vector as the input
nodes. 
Next, the system implements noise reduction in the data set
to improve the classification of user ratings. 
Finally, a deep belief
network and sentiment analysis (DBNSA) achieves data learning
for the recommendations

\begin{table*}[hbtp]
\centering
\caption{Properties of the considered methods are compared here.} 
\scalebox{0.8}{
\begin{tabular}{|c| c| c |c|}
\cline{1-4}
\multirow{1}{*}{\textbf{Models}} 
&\multirow{1}{*}{\textbf{Explicit Ratings}}
& \multirow{1}{*}{\textbf{Side information}}
&\multirow{1}{*} {\textbf{Model}}
\\ \cline{1-4}
SVD++ & \checkmark  & users' preferences generated from rating & SVD \\ 

ConvMF &\checkmark  &  users' review & Hybrid\\ 

VMCF & \checkmark  & product-images, also-viewed activity & SVD+Deep neural network\\ 

SBMF+R & \checkmark & ratings, reviews, and reliability of feedback & Probabilistic matrix factorization  \\ 

DMF & \checkmark  & degree of preference from users' ratings &  Hybrid \\

HCFDS & \checkmark  & user-product interaction & Hybrid  \\
DBNSA & \checkmark & users' review and product description & Deep belief network  \\

CM+RIM & \checkmark  & user-product interactions & Deep neural network \\

\cline{1-4}

FusionDeepMF &\checkmark & users' reliability & Fusion of MF and MLP \\
\cline{1-4}
\end{tabular} 
}
\label{tab:pro}
\end{table*}


%
\subsection{Evaluation Metrics}

%
The error evaluation metrics \textit{Root Mean Square Error (RMSE)} and \textit{Mean Absolute Error (MAE)} are applied to evaluate the performances of our proposed model.
We apply the ranking metrics $i.e.,$  \textit{Precision}, \textit{Recall}, \textit{F1-score} and \textit{Normalized Discounted Cumulative Gain (NDCG)} to evaluate product ranking  of  our models and baselines.

$RMSE$ is formulated as
 \begin{equation}\label{eq:MSE}
RMSE= \sqrt{\dfrac{\sum_{(i,j)\epsilon \tau} ( \hat {r_{ij}}- r_{ij})^{2} }{\mid\tau\mid}},
\end{equation}
and $MAE$ is formulated as
  \begin{equation}\label{eq:MAE1}
MAE= \dfrac{\sum_{(i,j)\epsilon \tau} \vert \hat {r_{ij}}- r_{ij}\mid }{\mid\tau\mid},
\end{equation}

where, $\mid\tau\mid$ presents our predicted ratings set. 

$Precision$, $Recall$, \textit{F1-score}, Mean Average Precision (MAP) and $NDCG$ are formulated as follows:

\begin{equation}\label{eq:M}
Precision= \dfrac{1}{n}{\sum_{i=1}^{n} \dfrac{\vert  Pre(i)\cap Orig(i)\mid }{\mid Pre (i)\mid}},
\end{equation}
\begin{equation}\label{eq:MAE2}
Recall= \dfrac{1}{n}{\sum_{i=1}^{n} \dfrac{\vert  Pre(i)\cap Orig(i)\mid }{\mid Orig (i)\mid}},
\end{equation}
\begin{equation}\label{eq:MAE3}
F1-score= \dfrac{2 \times Precision \times Recall}{Precision + Recall},
\end{equation}
\begin{equation}\label{eq:MAP}
MAP= \dfrac{1}{n}{\sum_{i=1}^{n} \dfrac{\sum_{t=1}^{y}Prec@t(i).I_{i,t} }{\mid Orig (i)\mid}},
\end{equation}
\begin{equation}\label{eq:MA}
NDCG= \dfrac{1}{n}\sum_{i=1}^{n}NDCG(i)=\dfrac{1}{n}\sum_{i=1}^{n}\dfrac{1}{Z_{i}}DCG(i),
\end{equation}

where, 

  \begin{equation}\label{eq:MAEE}
DCG(i)= {\sum_{j\in \omega(i)}\dfrac{2^{\hat{r}_{ij}}-1}{log_{2}(1+rank(j))}}.
\end{equation}

Here, $\omega(i)$ indicates the number of products in the test set   those are rated by $u_{i}$.
$Orig(i) = \{{j} \in \omega(i) \mid r_{ij} \geqslant 3\}$ indicates  the count of preferable products of  $u_{i}$, that is treated as ground truth $r_{ij}$ in  test dataset and 
$Pre(i) = \{{j} \in \omega(i) \mid \hat{r}_{ij} \geqslant 3\}$ indicates the count of suggested products to  $u_{i}$ based on rating prediction.
%
%
$rank(j)$ denotes the ranking place of $p_j$ in the sorted list of $\omega(i)$, based on  prediction of users' posted rating values.
Based on  the actual rating values in the test set, the normalized factor $Z_{i}$ is the DCG value of the original ranking place of  $\omega(i)$. 
The range of NDCG  is (0,1).
The  top-$t$ based ranking metric is denoted by \textit{Precision@t}.
\textit{Precision@t} is denoted by \textit{Prec@t}.
The length of ranking list of  $u_{i}$ for her buying products is denoted as $s$.
$I_{i,t}$ = 1 if the product at the $t^{th}$ place is defined by  $u_{i}$ and 0 otherwise.
%
%




\begin{table}[hbtp] 
\centering
\caption{Comparative RMSE results among \textit{MF}, \textit{MLP} and \textit{FusionDeepMF} on the Amazon.com online reviews dataset.}
%
\scalebox{0.8}{
\begin{tabular}{|c| c |  c c   |c| }
\cline{1-5}
\multirow{1}{*}{\textbf{Dataset}}
&\multirow{1}{*}{\textbf{\ Training size}}
&\multirow{1}{*} {\textbf{\ MF}}
&\multirow{1}{*}{\textbf{\ MLP}} 
&\multirow{1}{*} {\textbf{\ FusionDeepMF}}\\
\cline{1-5}

\multirow{4}{*} {Electronics}  & 40\%&1.663& 1.621& 1.593 \\

&50\% &1.411 &1.363 &1.201 \\

&60\% &1.311 & 1.263 & 1.022 \\

&70\% &1.271 & 1.153 &\textbf{0.907}  \\ \cline{1-5}

\multirow{4}{*} {Books}  & 40\%&1.373 & 1.401& 1.321\\

&50\% & 1.193& 1.223& 1.161  \\

&60\% &1.101 & 1.141 & 0.968  \\

&70\% & 1.051&  1.120&\textbf{0.827}  \\ \cline{1-5}

\multirow{4}{*} {Music}  & 40\%& 1.393& 1.411&1.355  \\

&50\% &1.196 & 1.261&1.101 \\

&60\% &0.991 & 1.057 & 0.852 \\

&70\% & 0.874& 0.927 &\textbf{0.724} \\ \cline{1-5}

\multirow{4}{*} {Movies \& TV}  & 40\%& 1.401&1.312 & 1.237 \\

&50\% &1.291 &1.146 &  1.093 \\

&60\% &1.011 & 0.962 &  0.830   \\

&70\% &0.901 & 0.861 &\textbf{0.711}\\ \cline{1-5}

\multirow{4}{*} {Home \& Kitchen}  & 40\%&1.577 & 1.543& 1.511 \\

&50\% & 1.391& 1.311& 1.253  \\

&60\% &1.292 &  1.123&  0.911  \\

&70\% &1.101 & 0.959 &\textbf{0.867}  \\ \cline{1-5}

\multirow{4}{*} {Amazon Instant Video}  & 40\%& 1.687&1.735 &1.637 \\

&50\%& 1.493 & 1.512&1.461  \\

&60\%& 1.071 &1.271  & 0.926 \\

&70\% &  0.996& 1.074&\textbf{0.896} \\ \cline{1-5}

\end{tabular} 
}
\label{tab:tuning}
\end{table}

\begin{table}[hbtp]
\centering
\caption{Comparison of error evaluation results on  all users among the baselines (present in $3^{rd}$ to $10^{th}$ column) and our approach (present in $11^{th}$  column).
Best performances are marked in bold.}
\scalebox{0.7}{
\begin{tabular}{|c| c|c   c cc c c c c |c| }
\cline{1-11}
\multirow{1}{*}{\textbf{Dataset}} 
&\multirow{1}{*}{\textbf{\ Metrics}}  
&\multirow{1}{*}{\textbf{\ SVD++}}
&\multirow{1}{*}{\textbf{\ ConvMF}}  
&\multirow{1}{*}{\textbf{\ VMCF}} 
&\multirow{1}{*} {\textbf{\ SBMF+R}}
&\multirow{1}{*} {\textbf{\ DMF}}
&\multirow{1}{*} {\textbf{\ HCFDS}}
&\multirow{1}{*} {\textbf{\ CM+RIM}}
&\multirow{1}{*} {\textbf{\ DBNSA}}

&\multirow{1}{*} {\textbf{\ \textit{FusionDeepMF}}}  \\ \cline{1-11}

\multirow{2}{*}{Electronics} 
&RMSE & 1.371 & 1.294& 1.121&1.107 &1.102&1.101&1.087& 1.067&\textbf{0.907}  \\

&MAE & 1.311& 1.207 &1.032 &0.981& 0.977&0.957&0.934&0.922  &\textbf{0.884}  \\\cline{1-11}

 \multirow{2}{*}{Books} & RMSE &1.167 & 1.101 & 0.927 & 0.921&0.911&0.901&0.891&0.879 &\textbf{0.827} \\ 

 & MAE & 0.974 & 0.912 & 0.871 &0.815 &0.812&0.809&0.791& 0.783 & \textbf{0.753} \\ \cline{1-11}

 \multirow{2}{*}{Music} &RMSE & 1.153 & 0.998 & 0.913 & 0.821 &0.811&0.809&0.796& 0.789  &\textbf{0.724} \\ 

 &MAE & 0.961 & 0.918 & 0.856 & 0.805 &0.803&0.801&0.781&0.766  &\textbf{0.710} \\ \cline{1-11}

\multirow{2}{*}{Movies \& TV} &RMSE & 1.147 & 0.987 & 0.901 &0.811 &0.802& 0.800&0.788&0.778&  \textbf{0.711}\\ 

&MAE & 0.946 & 0.911 & 0.861&0.773 &0.770&0.769 &0.759& 0.741 &  \textbf{0.701}\\ 
 \cline{1-11}

 \multirow{2}{*}{Home \& Kitchen}

 & RMSE &1.281 & 1.211 & 1.173 &1.023 &1.011&0.983&0.966& 0.947  &\textbf{0.867}\\

  & MAE & 1.175 & 1.031 & 0.985& 0.911&0.907&0.904 &0.891&0.861  &\textbf{0.798}\\\cline{1-11} 
\multirow{2}{*}{Amazon Instant Video}

 &RMSE  & 1.312&1.276 & 1.145 & 1.112&1.101&1.087&0.981& 0.967  & \textbf{0.896}\\
 &MAE & 1.216&1.122  & 1.011 & 0.991&0.988&0.977&0.951& 0.943 &  \textbf{0.838}\\
\cline{1-11}
\end{tabular} 
}

\label{tab:baselinecom}
\end{table}

\begin{table} [hbtp]
\centering
\caption{Performance of ranking metrics results between the baselines (present in $3^{rd}$ to $10^{th}$ column) and our model (present in $11^{th}$  column).}
\scalebox{0.7}{
\begin{tabular}{|c| c |  c c cc c c c c| c| }
\cline{1-11}
\multirow{1}{*}{\textbf{Dataset}}
&\multirow{1}{*}{\textbf{\ Metrics}}
 & \multirow{1}{*}{\textbf{\ SVD++}}
 &\multirow{1}{*} {\textbf{\ ConvMF}} 
 & \multirow{1}{*}{\textbf{\ VMCF}}
 &\multirow{1}{*} {\textbf{\ SBMF+R}}
 &\multirow{1}{*} {\textbf{\ DMF }}
 &\multirow{1}{*} {\textbf{\ HCFDS}}
 &\multirow{1}{*} {\textbf{\ CM+RIM}}
 &\multirow{1}{*} {\textbf{\ DBNSA }}
&\multirow{1}{*} {\textbf{\ \textit{FusionDeepMF}}}  \\ \cline{1-11}

\multirow{2}{*} {Electronics} 


  &F1-score  & 0.572 &0.608  & 0.650 & 0.723 &0.731&0.745 &0.772&0.781 & \textbf{0.820}   \\

   &NDCG &0.671 & 0.696& 0.707& 0.744&0.751 &0.771&0.782&0.793& \textbf{0.846} \\ \cline{1-11}

\multirow{2}{*}{Books}



  &F1-score   &0.580  & 0.617 & 0.663 &0.774 & 0.780&0.782 &0.791&0.801& \textbf{0.844}   \\

   &NDCG&0.644  & 0.656& 0.741&0.761 &0.783&0.787 &0.804&0.811 &  \textbf{0.881} \\ \cline{1-11}

 \multirow{2}{*}{Music} 


  &F1-score   & 0.628 & 0.667 &0.704  & 0.716 &0.733&0.741 &0.778&0.797  & \textbf{0.837}   \\

   &NDCG &0.658 & 0.704&0.728 &0.751 &0.773&0.789&0.803 &0.823& \textbf{0.883} \\ \cline{1-11}

 \multirow{2}{*}{Movies \& TV}


  &F1-score   & 0.617 & 0.688 & 0.729 & 0.744 &0.763&0.783&0.791& 0.805& \textbf{0.854}   \\

   &NDCG &0.638 &0.717 & 0.744& 0.763&0.780 &0.791&0.811& 0.845&  \textbf{0.890} \\ \cline{1-11}

\multirow{2}{*}{Home \& Kitchen} 


  &F1-score   & 0.619 & 0.664 & 0.698 & 0.712 &0.731&0.745 &0.762&0.776  & \textbf{0.815}   \\

   &NDCG &0.623 &0.673 & 0.703& 0.732&0.744&0.759 &0.775&0.784& \textbf{0.853} \\ \cline{1-11}
\multirow{2}{*} {Amazon Instant Video}  


  &F1-score & 0.651 & 0.694 &  0.713& 0.755& 0.769&0.783&0.795&0.811  & \textbf{0.862}   \\

   &NDCG& 0.679 & 0.704&0.730 & 0.766&0.779 &0.787&0.801&0.823&  \textbf{0.887} \\ \cline{1-11}
\end{tabular} 
}
\label{tab:baselinecomIR}
\end{table}














\subsection{Parameter Settings}

To scrutinize the capability of the \textit{MF} model in handling data sparsity problem,   we  set the latent dimension $K$ = 256 based on an extensive experimental observations.
%
%
%
%
We set $\lambda$ = 0.1 for Electronics, Home \& Kitchen, Amazon Instant Video datasets.
For the other three datasets we set $\lambda$ = 0.01. 
%
%


%
We implement \textit{FusionDeepMF} on the Keras.
We apply the mini-batch Adaptive Moment Estimation ($Adam$)  as an optimizer and Mean Absolute Error as the loss function in this model.
%
%
%
For neural network, we randomly initialize model parameters with a Gaussian distribution (with a mean of 0 and standard deviation of 0.01).
We  test the  layers of (2, 4, 6, 8), the batch size of (128, 256, 512, 1024) and the epochs of (1, 5, 10, 15, 20, 25, 30, 40, 50, 60). 
Our model's capability is determined based on the embedding size (predictive factors) of the last hidden layer.
We perform experiment on  the factors of (8, 16, 32, 64, 128).
We also perform on   larger factors but large factors may cause overfitting and degrade  the performance.
After observation, we choose 4 hidden layers, 512 batch size  and 12 epochs  in our deep learning model.
For example, if the size of predictive factors is 64, then architecture of the neural $CF$ layers is 512 $->$ 256 $->$ 128 $->$ 64, and embedding size is 256.
Details of observation  are discussed later.
For the \textit{FusionDeepMF} with pre-training,
$\gamma$ was set to 0.5, allowing the pre-trained \textit{MF} and \textit{MLP} to
contribute equally to \textit{FusionDeepMF} initialization.
In our experiment, the respective optimal parameters are chosen based on our investigations.
Based on the validation dataset, we tune the hyper-parameter.
\subsection{Performance Analysis}

To evaluate the performances  of the \textit{FusionDeepMF}, \textit{MF} (only linear) and \textit{MLP} (only non-linear)  on our experimental dataset with different training size,
five-fold cross validation is performed.
For training purpose $x$\% of  data, for validation purpose ($\dfrac{100-x}{2}$)\% of data are used  and  the remaining data is used for testing phase.
Table~\ref{tab:tuning} follows the above settings.
%
For all the other experiments throughout the paper, we follow the following setting:
To evaluate hyper-parameter of  our model,  five-fold cross validation is performed.
In each fold,  for training 70\% of  data, for validation 15\% of data  and  for testing purpose the remaining 15\% of data are set in both our approach and baselines.
%
%
We perform random sampling  five times and  calculate the mean as the final output for each. 
%
%

Table~\ref{tab:tuning} shows $RMSE$ results from the  \textit{MF}, \textit{MLP} and fusion of \textit{MF} and \textit{MLP} \textit{FusionDeepMF} based on the data.
For all the datasets, the \textit{FusionDeepMF} performs better compared to the \textit{MF} model and \textit{MLP} model.
The two separate embeddings, generated by the \textit{MF} and \textit{MLP} model simultaneously, can mutually reinforcement each other to learn better rating value.
This is the advantage of the fusion of the \textit{MF} and \textit{MLP}.
If we compare the performance between the \textit{MF} and \textit{MLP}, Table~\ref{tab:tuning} shows that the \textit{MLP} performs better on Electronics, Movies \& TV and Home \& Kitchen datasets.
The linear kernel of \textit{MF} model performs better on Books, Music and Amazon Instant Video datasets.
For these three datasets, the linear kernel learns users' features very accurately compared to non-linear kernel and improves overall performances.
For this reason, the idea of the fusion of \textit{MF} and \textit{MLP} is very effective.


The comparison of $RMSE$, $MAE$ results of predicted rating and original rating between the baseline models and our model based on the Amazon.com online review dataset is shown in Table \ref{tab:baselinecom}.
In this table,
$3^{rd}$ to $10^{th}$ column present the performance of baseline models.
The last  column shows  our model's performances.
For each category data, our model performs better significantly.
%
%
The significant performance of our model indicates the importance of fusion between linear and non-linear kernel and the importance of reliability score.  
Our approach achieves  better performances than $DBNSA$ which indicates the effectiveness to learn users' and products' dual features embeddings based on  users' reliability score through our fusion model.
In $SBMF+R$ model the authors have considered users' review text and reliability score based on helpful votes.
But this reliability concept is different from our reliability concept and we learn this features using linear and non-linear kernel based methods where $SBMF+R$ model is based on probabilistic matrix factorization method.
For the fusion between linearity and non-linearity our model learns users' features very accurately.

$DMF, HCFDS$ and $CM+RIM$ are developed based on user-product interaction ratings and these three baselines use deep learning model to learn users' features, where our model learns users' features based on reliability score using linear and non-linear kernel.
For this  learning technique our model outperforms them.
Our approach also achieves   better performances  over the \textit{ConvMF} and \textit{VMCF} for its dual embedding features learning capability.
In Table \ref{tab:baselinecomIR}, the evaluation of IR metrics  between the baselines and our approach  are presented.
In this table,
$3^{rd}$  to $10^{th}$ column presents the performance of the baselines.
The last  column shows the performance of our model.
For each category data, our model \textit{FusionDeepMF}  performs better significantly.


%
In section~\ref{timec}, we have discussed the time complexity of our model.
In Table~\ref{tab:execution}, we present the training time of the baseline models and our model.
We  conduct this experiment on a Intel Core 3.07 GHz CPU and 128 GB RAM with Ubuntu 16.04.
Our approach is more effective compared to $DMF$, $VMCF$, $CM+RIM$ and $ConvMF$ based on the total empirical runtime  in the training phase.
The training sample of  $VMCF$  contains one observed rating and two side information as product images and also-viewed activity.
$VMCF$ applies pre-trained $CNN$ to extract features from the pictures of products.
Then \textit{PMF} based approach is applied to understand users' rating activities  from extracted features of product pictures and also-viewed information of users.
Therefore, latent factor matrices of the users and
products  have to be updated twice for one training phase. 
The empirical training time  of $VMCF$ for each iteration is lower than our methodology.
But $VMCF$ converges approximately after 15 iterations, where our model converges after 12 iterations.
For this reason the empirical total runtime of  our methodology is slightly better than  $VMCF$.
The time complexity
of $ConvMF$ is $\mathcal{O}$($k^{2}n_{R}$ + $k^{3}N$ + $k^{3}M$ + $n_{c} . p . l . M$)~\cite{kim2016convolutional}. 
In this hybrid methodology, linear and non-linear kernel are integrated.
$CNN$ is applied to analyze both review text and Gaussian noise.
Then the objective function is used to optimize variables using users' posted rating values and contextual data.
$ConvMF$ converges approximately after 16 iterations.
Due to analyzing review text, its empirical runtime is more than our approach.

If we analyze training times of $DMF$ and $CM+RIM$, we can find that empirical runtime of each iteration of these models is lower than our model, but total train time is higher than our model because
our model converges approximately after 12 iterations, where these two baselines converge  approximately  after 21 iterations.
This indicates the effectiveness of the learning technique with reliability score.
$SVD++$ training sample contains
only users' posted rating values.
 $SVD++$ needs to update  the latent factor matrices of the users and products  once for each training sample, where our methodology learns users' features using linear and non-linear way.
For this reason the empirical runtime of  our methodology is slightly higher than  $SVD++$.
Our model converges approximately after 12 iterations, where these two baseline $HCFDS$ and  $DBNSA$ converge  approximately  after 16 or 17 iterations.
If we analyze the train times of  $HCFDS$ and  $DBNSA$, we can find that empirical runtime of each iteration of these models is lower than our model, but the total training time is higher than our model.
The empirical runtime of our model is almost similar to $SBMF+R$, but according to the recommendation accuracy our model is the best.

%

















\begin{table}[hbtp]
\centering
\caption{Comparison on  total time needed (in second) to learn the parameters   between baselines (present in $2^{nd}$ to $9^{th}$ column) and our approach (present in $10^{th}$  column).}
\scalebox{0.7}{
\begin{tabular}{|c| c   c cc c c c c |c| }
\cline{1-10}
\multirow{1}{*}{\textbf{Dataset}}   
&\multirow{1}{*}{\textbf{\ SVD++}}
&\multirow{1}{*}{\textbf{\ VMCF}} 
&\multirow{1}{*}{\textbf{\ ConvMF}} 
&\multirow{1}{*} {\textbf{\ DMF}}
&\multirow{1}{*} {\textbf{\ HCFDS}}
&\multirow{1}{*} {\textbf{\ CM+RIM}}
&\multirow{1}{*} {\textbf{\ SBMF+R}}
&\multirow{1}{*} {\textbf{\ DBNSA}}

&\multirow{1}{*} {\textbf{\ \textit{FusionDeepMF}}}  \\ \cline{1-10}

\multirow{1}{*}{Electronics} 
 &825  &1215 & 1987& 1688 & 1137&1632&1033 &1175& 1030  \\
 
 \multirow{1}{*}{Books} 
 & 912 & 1257& 2216&1697 & 1325&1641 &1136 &1320& 1143 \\
 
  \multirow{1}{*}{Music} 
 & 883 & 1085& 2198& 1560 & 1025&1549&981 &1033 & 995\\

\multirow{1}{*}{Movies \& TV} 
 & 871 & 1011& 2011&  1331& 980&1357 & 903& 997& 911\\

 \multirow{1}{*}{Home \& Kitchen} 
 & 851 & 1307& 2083& 1463 & 1201& 1492& 1113& 1229&1107 \\
 
 \multirow{1}{*}{Amazon Instant Video} 
 & 837 & 1273&2077 & 1366 & 1157& 1384& 1012& 1163&1009 \\
\cline{1-10}

\end{tabular} 
}

\label{tab:execution}
\end{table}

\subsection{Utility of Pre-training}

To understand  the effectiveness of pre-training of \textit{FusionDeepMF}, we observe the performance of two versions of our model with and without pre-training. 
As observed in Table~\ref{tab:pretraining}, the \textit{FusionDeepMF} with pre-raining performs better.
This result proves  the effectiveness of the pre-training strategy in our model.
%
\begin{table}[hbtp]
\centering
\caption{Performance of MAE   of \textit{FusionDeepMF} for rating prediction with and without pre-training.}
\scalebox{0.9}{
\begin{tabular}{|c|  c | c|}
\cline{1-3}
\multirow{1}{*}{\textbf{Dataset}}
&\multirow{1}{*}{\textbf{with pre-training}}
& \multirow{1}{*}{\textbf{without pre-training}} \\ \cline{1-3}

\multirow{1}{*}{Electronics}


 &\textbf{0.884} &0.911  \\ 

\multirow{1}{*}{Books} 


 & \textbf{0.753} &0.778     \\ 

\multirow{1}{*}{Music} 


 &\textbf{0.710} &0.730   \\ 

\multirow{1}{*}{Movies \& TV}


 & \textbf{0.701}  &0.729  \\ 

 \multirow{1}{*}{Home \& Kitchen} 
 

 &\textbf{0.798}  &0.835  \\ 

 \multirow{1}{*}{Amazon Instant Video}
 

 &\textbf{0.838}  &0.855    \\ \cline{1-3}
 
 \end{tabular}
}
\label{tab:pretraining}
\end{table}
 \begin{table}[hbtp]
\centering
\caption{Results for our method with different predictive factors of the final latent space.}
\scalebox{0.86}{
\begin{tabular}{|c| c| c c c c c|}
\cline{1-7}
\multirow{1}{*}{\textbf{Dataset}}
&\multirow{1}{*}{\textbf{Metric}}
&\multirow{1}{*}{\textbf{$8$}}
& \multirow{1}{*}{\textbf{$16$}}
&\multirow{1}{*} {\textbf{$32$}}  
&\multirow{1}{*} {\textbf{$64$}}
&\multirow{1}{*} {\textbf{$128$}}\\ \cline{1-7}

\multirow{2}{*}{Electronics} 

& RMSE& 1.203& 1.137& 1.011& 0.907 & \textbf{0.889}\\ \cline{2-7}
& MAE&1.171 &1.012 & 0.975& 0.884&\textbf{0.879}
\\ 
\cline{1-7}

\multirow{2}{*}{Books} 

& RMSE& 1.193& 1.044& 0.943& \textbf{0.827} &0.833\\ \cline{2-7}
& MAE& 1.047 & 0.955&0.861& \textbf{0.753}&0.766 \\ 
\cline{1-7}

\multirow{2}{*}{Music}
& RMSE& 1.084& 0.971&0.844& \textbf{0.724}  &0.738\\ \cline{2-7}
& MAE& 1.007&0.947 & 0.817& \textbf{0.710}& 0.719\\ \cline{1-7}

\multirow{2}{*}{Movies \& TV}
& RMSE& 0.993& 0.875& 0.796& \textbf{0.711} &0.727\\ \cline{2-7}
& MAE& 0.894& 0.801& 0.758&\textbf{0.701}& 0.719\\ \cline{1-7}

\multirow{2}{*}{Home \& Kitchen}
& RMSE&1.177 & 0.991&0.901&  0.867 &\textbf{0.851}\\ \cline{2-7}
& MAE&0.945 &0.881 &0.837 & 0.798& \textbf{0.779}\\ \cline{1-7}

\multirow{2}{*}{Amazon Instant Video}
& RMSE&  1.169& 0.989&  0.907&0.896&\textbf{0.873}\\ \cline{2-7}
& MAE& 1.032& 0.965& 0.918& 0.838&\textbf{0.820} \\ \cline{1-7}
\end{tabular} 
}
\label{tab:Factor}
\end{table}
\begin{table}[hbtp]
\centering
\caption{Performance of MAE  with different number of layers of \textit{FusionDeepMF} for rating prediction.}
\scalebox{0.8}{
\begin{tabular}{|c|  c c  c c|}
\cline{1-5}
\multirow{1}{*}{\textbf{Dataset}}
&\multirow{1}{*}{\textbf{MLP-2}}
& \multirow{1}{*}{\textbf{MLP-4}}
&\multirow{1}{*} {\textbf{MLP-6}}  
&\multirow{1}{*} {\textbf{MLP-8}} \\ \cline{1-5}

\multirow{1}{*}{Electronics}


 &0.911&\textbf{0.884} &0.889& 0.894 \\ 

\multirow{1}{*}{Books} 


 &0.781 & \textbf{0.753} &0.761  &0.767  \\ 

\multirow{1}{*}{Music} 


 &0.733 &0.710 &\textbf{0.701} &0.716 \\ 

\multirow{1}{*}{Movies \& TV}


 &0.729 & \textbf{0.701} &0.711 &0.719 \\ 

 \multirow{1}{*}{Home \& Kitchen} 
 

 &0.821 &0.798 &\textbf{0.789} &0.801  \\ 

 \multirow{1}{*}{Amazon Instant Video}
 

 &0.857 &\textbf{0.838} &0.843& 0.852 \\ \cline{1-5}

\end{tabular}
}
\label{tab:layers}
\end{table}

\subsection{Different settings of predictive factors}
The sensitivity of the factors in each layer is high  in our approach.
The performances  with different number of factors on the top final latent space are compared. 
The investigation on our method with four layer are arranged and we set the number of the total factors on the top layer from 8 to 128. 
In Table~\ref{tab:Factor}, it is shown that the final layer with 64 factors achieves the best performance on Books, Music and Movies \& TV datasets.
For the other three datasets, the final layer with 128 factors gets the best performance.
From this observation we can say that the final representations with more predictive factors might be more effective when data is small and sparse in nature.

\subsection{Different Number of Layers}

We  investigate whether deep neural network is suitable for our recommendation system or not.
We experiment with different number of hidden layers
and Table~\ref{tab:layers} shows the results.
In this table,
 $2^{nd}$ to $5^{th}$ column show the performance of the \textit{FusionDeepMF} model with different number of hidden layers. 
\textit{MLP}-$2$ means two hidden layers are used in the \textit{FusionDeepMF} model.
In most of the cases, we are getting good performance when four hidden layers are used.
As shown in Table \ref{tab:layers}, on Electronics, Books, Movies \& TV, Amazon Instant Video datasets our model with layer 4 illustrates the best performance.
For the other two datasets Music and Home \& Kitchen, our model with layer 6 illustrates the best performance.

We also observed the similar  performance of our approach with different number of layers when $RMSE$ is applied.
From this observation, we can say that using a deep network structure with stacking more hidden
layers  is not always beneficial for a  recommendation task.

\subsection {Ranking of Top-$t$ products recommendation}

Evaluation of   ranking metrics@Top-$t$ between the baselines and our approach are represented in Table \ref{tab:rank}.
The $1^{st}$ column of this table presents different categories of data,
$3^{rd}$ to $10^{th}$ column show the performance of the baselines.
The last column shows the evaluation of our approach.
For all category data, \textit{FusionDeepMF} performs the best.

\begin{table} [hbtp]
\centering
\caption{Comparison with  baselines (present in $3^{nd}$ to $10^{th}$ column) and our methodology (present in last  column) based on  Top-$t$ products recommendation.}
\scalebox{0.66}{
\begin{tabular}{|c| c | c c c c c c c c|c| }
\cline{1-11}
\multirow{1}{*}{\textbf{Dataset}}
&\multirow{1}{*}{\textbf{\ Metrics}}
&\multirow{1}{*}{\textbf{\ SVD++}} 
& \multirow{1}{*}{\textbf{\ ConvMF}}
&\multirow{1}{*} {\textbf{\ VMCF}}
&\multirow{1}{*} {\textbf{\ SBMF+R}}
& \multirow{1}{*}{\textbf{\ DMF}} 
& \multirow{1}{*}{\textbf{\ HCFDS}} 
&\multirow{1}{*} {\textbf{\ CM+RIM}}
 &\multirow{1}{*} {\textbf{\ DBNSA }}
&\multirow{1}{*} {\textbf{\ FusionDeepMF}}  \\ \cline{1-11}
\multirow{4}{*} {Electronics} 

& NDCG & 0.671 &0.696  &0.707 & 0.744 &0.751&0.771 &0.782&0.793  & \textbf{0.846}\\ 

&MAP& 0.602 &0.621  &0.688 & 0.732 &0.740&0.745 &0.765&0.771  & \textbf{0.828}  \\

&F1@5& 0.497 & 0.557 &0.597 &0.601 &0.609&0.617 & 0.628&0.643 & \textbf{0.733}  \\

&F1@10& 0.511 & 0.587 &0.604 &0.611 &0.620&0.634 &0.641 &0.655 & \textbf{0.747} \\

\cline{1-11}

\multirow{4}{*}{Books} 

&NDCG&0.644  & 0.656& 0.741&0.761 &0.783&0.787 &0.804&0.811 &  \textbf{0.881} \\

&MAP& 0.599 &0.611  &0.688 & 0.727 &0.740&0.751 &0.771&0.782  & \textbf{0.837} \\

&F1@5& 0.498 & 0.587 &0.647 &0.649 &0.653&0.658 & 0.660&0.662 & \textbf{0.771}  \\

&F1@10& 0.511 & 0.603 &0.651 &0.657 &0.661&0.667 & 0.670&0.675 & \textbf{0.784} \\
\cline{1-11}

\multirow{4}{*}{Music} 

&NDCG &0.658 & 0.704&0.728 &0.751 &0.773&0.789&0.803& 0.823& \textbf{0.883} \\

&MAP& 0.602 &0.689  &0.701 & 0.733 &0.750&0.767 &0.778&0.791  & \textbf{0.831} \\

&F1@5& 0.576 & 0.604 &0.669 &0.714 &0.721&0.727 &0.733& 0.742 & \textbf{0.778}  \\

&F1@10& 0.598 & 0.622 &0.689 &0.722 &0.728&0.731 & 0.639&0.750 & \textbf{0.782} \\
\cline{1-11}

\multirow{4}{*}{Movies \& TV} 

   &NDCG &0.638 &0.717& 0.728& 0.763&0.780& 0.791&0.811& 0.845&  \textbf{0.890} \\

&MAP& 0.605 &0.694  &0.709 & 0.729 &0.739&0.753 &0.785&0.801  & \textbf{0.844} \\

&F1@5& 0.567 & 0.600 &0.697 &0.711 &0.720&0.727 & 0.732&0.740 & \textbf{0.779}  \\

&F1@10& 0.587 & 0.613 &0.702 &0.717 &0.724&0.731 & 0.739&0.747 & \textbf{0.784} \\
\cline{1-11}

\multirow{4}{*}{Home \& Kitchen} 

&NDCG &0.623 &0.673 & 0.703& 0.732&0.744&0.759 &0.775&0.784& \textbf{0.853} \\

&MAP& 0.598 &0.621  &0.694 & 0.714 &0.730&0.743 &0.761&0.775  & \textbf{0.822} \\

&F1@5& 0.510 & 0.566 &0.608 &0.609 &0.622&0.634 &0.642 &0.647 & \textbf{0.701}  \\

&F1@10& 0.568 & 0.602 &0.633 &0.644 &0.661&0.671 & 0.676&0.680 & \textbf{0.717} \\
\cline{1-11}

\multirow{4}{*}{Amazon Instant Video} 

&NDCG& 0.679 & 0.704&0.730 & 0.766&0.779 &0.787&0.801&0.823&  \textbf{0.887} \\

&MAP& 0.665 &0.699  &0.713 & 0.751 &0.763&0.774 &0.791&0.804  & \textbf{0.853} \\

&F1@5& 0.557 & 0.681 &0.720 &0.722 &0.731&0.737 &0.741 &0.747 & \textbf{0.779}  \\

&F1@10& 0.576 & 0.694 &0.739 &0.741 &0.743&0.749 & 0.753&0.759 & \textbf{0.785} \\
\cline{1-11}




\end{tabular} 
}
\label{tab:rank}
\end{table}

\section{Conclusion and Future Work}\label{sec:6}

In this work, 
we propose the \textit{FusionDeepMF}  model, where  \textit{MF}  integrates a linear kernel to model the latent feature vectors based on users' rating activity and reliability score  and on the other hand \textit{MLP}  uses a non-linear kernel to learn users' latent feature vectors. 
This model uses a tuning-parameter determining the trade-off between the dual embeddings that are generated from the  linear and non-linear kernels.
We observe that the new way of modeling users' reliability score  not only characterizes the users' latent feature vectors accurately but also presents a better interpretation of how the similarity of users' reliability score affects rating prediction.
The raters also play a vital role to notify the reliability of a reviewer's review and help the model to learn the latent factor of the reviewer more accurately to predict rating.

%
%

%
We would like to further improve the performances of  our approach and would like to experiment on other  datasets.
There are many existing e-commerce companies, which  have links  with social media where users share their  opinion on their social pages.
We would like to investigate if social network feedback can be merged with the reliability score (review network feedback) to learn users' latent feature vectors more accurately.
Many online e-commerce companies  share users' posted feedback on their social sites.
%
%
We would prefer to analyze how social networks can be used to learn users' preferable zones.

We would prefer to experiment on  an attention layer based deep learning model to extract users' preferable products' features.
Some micro behaviors activities such as clicking activities, searching activity, add to cart, order, dwell time, etc. will also be considered in our model.
In~\cite{park2017also}, the authors are considering  ``also-viewed" categories for quality vector embedding.
We will consider on other relations among products, like ``also-bought", ``frequently-bought-together" and  ``bought-after-buying."
It has been decided to check whether the above relationships are effective in projecting good quality product vector embedding and  how to exactly suit  in our methodology.
This concept gives us other fascinating way to extend our approach including more characteristics of users.

\bibliographystyle{spmpsci}

\bibliography{www}
\end{document}